\documentstyle[12pt,bezier]{article}
\begin{document}
% ***************    NEW COMMANDS   *******************
\def \inbar{\vrule height1.5ex width.4pt depth0pt}
\def \xC{\relax\hbox{\kern.25em$\inbar\kern-.3em{\rm C}$}}
\def \xR{\relax{\rm I\kern-.18em R}}
\newcommand{\R}{\xR}
\newcommand{\C}{\xC}
\newcommand{\xZ}{Z \hspace{-.08in}Z}
\newcommand{\xbe}{\begin{equation}}
\newcommand{\be}{\begin{equation}}
\newcommand{\xee}{\end{equation}}
\newcommand{\ee}{\end{equation}}
\newcommand{\xbea}{\begin{eqnarray}}
\newcommand{\bea}{\begin{eqnarray}}
\newcommand{\xeea}{\end{eqnarray}}
\newcommand{\eea}{\end{eqnarray}}
\newcommand{\xnn}{\nonumber}
\newcommand{\nn}{\nonumber}
\newcommand{\xkt}{\rangle}
\newcommand{\kt}{\rangle}
\newcommand{\xbr}{\langle}
\newcommand{\br}{\langle}
\newcommand{\xcun}{\mbox{\footnotesize${\cal N}$}}
\newcommand{\cun}{\mbox{\footnotesize${\cal N}$}}
\newcommand{\cum}{\mbox{\footnotesize${\cal M}$}}
\title{Perturbative Calculation of  the Adiabatic Geometric
Phase and Particle in a Well with Moving Walls}
\author{Ali Mostafazadeh\thanks{E-mail address: 
amostafazadeh@ku.edu.tr}\\ \\
Department of Mathematics, Ko\c{c} University,\\
Istinye 80860, Istanbul, TURKEY}
\date{ }
\maketitle

\begin{abstract}
We use the Rayleigh-Schr\"odinger perturbation theory to calculate the 
corrections to the adiabatic geometric phase due to a perturbation of the 
Hamiltonian. We show that these corrections are at least of second order in
the perturbation parameter. As an application of our general results we
address the problem of the adiabatic geometric phase for a one-dimensional 
particle which is confined to an infinite square well with moving walls. 
\end{abstract}
%\vspace{2mm}

%PACS numbers: 03.65.Bz
%\vspace{2mm}

\baselineskip=24pt

\section{Introduction} 

In Ref.~\cite{pe-pr}, Pereshogin and Pronin consider the problem of the 
calculation of the adiabatic geometric phase \cite{berry1984} for a free
particle confined between moving walls. The quantum dynamics of
this system has been studied by Doescher and Rice \cite{do-ri}, Munier
{\em et al} \cite{mu-bu-fe-fi}, Berry and Klein \cite{be-kl},
Greenbereger \cite{greenberger}, Pinder \cite{pinder}, Seba \cite{seba},
Makowski {\em et al} \cite{ma-de}, Devoto and Pomorisac \cite{de-po} and 
Dodonov {\em et al} \cite{do-kl-ni}. The analysis of Pereshogin and Pronin 
\cite{pe-pr} is however different in nature, for it uses the geometric 
ideas of parallel transportation in vector bundles to derive an effective 
Hamiltonian for the system. They suggest that the quantum dynamics of the 
system is determined by the effective Hamiltonian and employ the 
Rayleigh-Schr\"odinger perturbation theory to obtain the first nonvanishing 
contribution to Berry's connection one-form (vector potential) for the 
effective Hamiltonian.

The phenomenon of the geometric phase induced by moving boundaries
was initially considered by Levy-Leblond \cite{le-le} who calculated the
phase shift of the wave function of a free particle which is forced to pass
through a wave guide of finite length. There is also a mention of a
`geometric phase' in Greenberger's analysis of the dynamics of a particle 
confined between moving walls \cite{greenberger}. Greenberger's terminology 
is however not appropriate, for what he calls a geometric 
phase depends on spatial coordinates. Therefore, it is not really the phase
of the state vector in the Hilbert space and must not be confused with the
geometric phase of Berry \cite{berry1984} and its nonadiabatic
generalization due to Aharonov and Anandan \cite{aa}. In fact, to the best
of the author's knowledge, Pereshogin and Pronin's article \cite{pe-pr} is
the only publication in which the authors use Berry's framework to study
the problem of the adiabatic geometric phase due to moving boundaries.

In the present article we address the problem of the perturbative calculation
of Berry's connection one-form for a general nondegenerate Hamiltonian.
Furthermore, we use the method of time-dependent quantum canonical 
transformations \cite{jmp97b,jpa98a} to study the dynamics of a particle 
confined between moving walls. We then apply our general results to obtain 
a perturbative expression for the adiabatic geometric phase for this
system. In particular, we shall consider the special case where the particle
is free and compare our results with those of Pereshogin and Pronin. 

The organization of the paper is as follows. In sections~2 and~3, we shall 
offer a brief review of the (cyclic and noncyclic) adiabatic geometric
phases and the Rayleigh-Schr\"odinger perturbation theory, respectively. In 
section~4, we derive an expression for the Berry's connection one-form 
which yields the perturbative corrections to the connection one-form for 
the nonperturbed system to arbitrary orders of perturbation. In section~5, 
we treat the quantum dynamics of a particle confined between moving 
walls. In section~6, we address the problem of the adiabatic geometric
phase for this system. In section~7, we summarize our main results
and conclude the paper with our final remarks.

\section{Adiabatic Geometric Phase}

Consider a parametric Hamiltonian $H[R]$ satisfying the following
conditions:
	\begin{itemize}	
	\item
$H[R]$ depends on a set of real parameters $R=(R^1,R^2,\cdots,R^d)$ 
which are identified with local coordinates of a smooth parameter 
manifold\footnote{Here $R$ abbreviates $(R^1,R^2,\cdots,R^d)$. This
notation does not mean that $R$ is a vector belonging to $\R^d$. $R$ is
a $d$-tuple of real numbers representing the coordinates of a smooth
parameter manifold. The latter must not be confused with the configuration
space of the corresponding system.};
	\item
$H[R]$ is a Hermitian operator with a discrete spectrum for all possible values 
of  $R$;
	\item
The eigenvalues $E_n[R]$ of $H[R]$ are nondegenerate for all possible values
of $R$. In particular as $R$ changes in time, no level-crossings occur.
	\item
The eigenvalues $E_n[R]$ of $H[R]$ are smooth functions of $R$.
	\end{itemize}

Now if the parameters $R$ change in time $t\in[0,\tau]$ in such a way that
the evolution of the system is adiabatic \cite{adiabatic,pra97a}, then a 
normalized eigenvector $|n;R(0)\kt$ of the initial Hamiltonian $H[R(0)]$
evolves according to \cite{berry1984}
	\be
	|\psi(t)\kt=e^{i\alpha_n(t)}|n;R(t)\kt\;, 
	\label{e1-1}
	\ee
where $\alpha_n(t)$ is a phase angle and $|n;R\kt$ is a normalized 
eigenvector of $H[R]$ corresponding to the eigenvalue $E_n[R]$, i.e., 
$|n;R\kt$ is a solution of
	\be
	H[R]|n;R\kt=E_n[R]|n;R\kt\;.
	\label{ei-va-eq}
	\ee
We shall assume that $|n;R\kt$ are smooth functions of $R$ and that they 
form a complete orthonormal set of basis vectors for the Hilbert space. This 
means that for all possible values of $R,~m,$ and $n$,
	\be
	\br m;R|n;R\kt=\delta_{mn}\;,~~~{\rm and}~~~
	\sum_n|n;R\kt\br n;R|=1\;.
	\label{ortho}
	\ee
The phase angle $\alpha_n(t)$ appearing in (\ref{e1-1}) is given by
	\be
	\alpha_n(t):=\delta_n(t)+\gamma_n(t)\;,
	\label{e1-2}
	\ee
where
	\bea
	\delta_n(t)&:=&-\frac{1}{\hbar}\int_0^tE_n(t')dt'\;,
	\label{e1-3}\\
	\gamma_n(t)&:=&\int_0^t
	{\cal A}_n(t')dt'=\int_{R(0)}^{R(t)}A_n[R]\;,
	\label{e1-4}\\
	{\cal A}_n(t)&:=&i\br n;R(t)|\frac{d}{dt}|n;R(t)\kt\;,~~~{\rm and}
	\label{e1-5}\\
	A_n[R]&:=&i\br n;R|d|n;R\kt=\sum_{a=1}^d i \br n;R|
	\frac{\partial}{\partial R^a}|n;R\kt dR^a\;.
	\label{e1-6}
	\eea
$\delta_n(t)$ and $\gamma_n(t)$ are called the {\em dynamical}
and {\em geometrical} parts of the {\em total phase angle} $\alpha_n(t)$,
respectively. The one-form $A_n[R]$ is known as {\em Berry's connection
one-form} \cite{berry1984}.

The adiabatic geometric phase \cite{p31,po-sj} is given by
	\be
	\Phi_n(t)=W_n(t) \Gamma^n(t) \;,
	\label{e1-7}
	\ee
where 
	\bea
	\Gamma^n(t)&:=&e^{i\gamma_n(t)}\;,	~~~{\rm and}
	\label{e1-7.1}\\
	W_n(t)&:=&\br n;R(0)|n;R(t)\kt\;.
	\label{e1-8}
	\eea
If the parameters $R(t)$ trace a closed path $C$ in the parameter space,
i.e., there is $T\in\R^+$ such that $R(T)=R(0)$, then at $t=T$ we have 
$H[R(T)]=H[R(0)]$, $|n;R(T)\kt=|n;R(0)\kt$, and $W_n(T)=1$. In 
particular, $|\psi(0)\kt=|n;R(0)\kt$ undergoes a cyclic evolution and 
$\Phi_n(T)$ yields the cyclic adiabatic geometric phase or the Berry's phase
\cite{berry1984}:
	\be
	\Phi_n(T)=\Gamma^n(T)=e^{i\gamma_n(T)}=e^{i\oint_C A_n[R]}\;.
	\label{e1-9}
	\ee

The above derivation of the geometric phase is valid even for the cases
where the Hilbert space is time-dependent. The time-dependence
of the Hilbert space may be reflected in the definition of the measure used
to define integration. For example consider the problem of a
one-dimensional particle confined between two walls positioned 
as $x=0$ and $x=L(t)$, where $L:[0,\tau]\to\R^+$ is a smooth function
and $\tau$ is the duration of the evolution of the system. The Hilbert
space ${\cal H}_t$ is $L^2([0,L(t)])$ which depends on time. However,
we can identify $L^2([0,L(t)])$  with
	\be
	L_\mu^2(\R):=\{\psi:\R\to\C~|~\int_{-\infty}^\infty
	\psi^*(x)\psi(x)\mu_t(x)dx<\infty\}\;.
	\label{L2-mu}
	\ee
where the measure function $\mu_t$ is given by
	\be
	\mu_t(x):=\theta(x)-\theta(x-L(t))\;,
	\label{mu}
	\ee
and $\theta:\R\to\{0,1\}$ denotes the step function:
	\be
	\theta(x):=\left\{\begin{array}{ccc}
	1&{\rm for}& x\geq 0\\
	0&{\rm for}&x<0\;.\end{array}\right.
	\label{e8}
	\ee

Now let us denote the eigenfunctions of $H[R(t)]$ in the position 
representation by $\phi_n$, i.e., $\phi_n(x;t):=\br x|n;R(t)\kt$. Then
	\bea
	{\cal A}_n^*&=&-i\left(\br n;R(t)|\frac{d}{dt}|n;R(t)\kt\right)^*
	=-i\left( \int_{-\infty}^{\infty}\phi_n^*\dot\phi_n\mu dx\right)^*
	=-i \int_{-\infty}^{\infty}\phi_n\dot\phi_n^*\mu dx\;,\nn \\
	&=&-i\frac{d}{dt}\int_{-\infty}^{\infty}\phi_n\phi_n^*\mu dx
	+i\int_{-\infty}^{\infty}\phi_n^*\dot\phi_n\mu dx+
	i\int_{-\infty}^{\infty}|\phi_n|^2\dot\mu dx\;,\nn\\
	&=&{\cal A}_n+i|\phi_n(x=L(t);t)|^2\;,
	\label{reality}
	\eea
where $^*$ stands for the operation of complex conjugation. The last
equation in (\ref{reality}) is obtained by making use of the fact that $\phi_n$
are normalized and that $\dot\mu=-\dot\theta(x-L(t))=\delta(x-L(t))$ where 
$\delta(x)$ is the Dirac delta function. Eq.~(\ref{reality}) shows that ${\cal A}_n$ 
and consequently the Berry connection one-form $A_n$ and the phase 
angles $\gamma_n(t)$ and $\alpha_n(t)$ are real, provided that one
chooses the boundary condition:  $\left.\phi_n\right|_{x=L(t)}=0$. 

\section{Perturbation Theory}

In order to compute the adiabatic geometric phase (\ref{e1-7}), one needs
to obtain the eigenvectors $|n;R\kt$ of the Hamiltonian $H[R]$. There are,
however, quite a few Hamiltonians whose eigenvalue equation is solved 
exactly. Often, one uses approximation schemes to obtain the eigenvalues 
and eigenvectors of a given Hamiltonian. One of the best-known 
approximation methods of solving the eigenvalue problem is the 
Rayleigh-Schr\"odinger perturbation theory \cite{dbohm-qm,abohm-qm}.
We shall next derive the basic results of the Rayleigh-Schr\"odinger 
perturbation theory.

Consider a parametric Hamiltonian of the form
	\be
	H[R]=H_0[R]+\epsilon[R] h[R]\;,
	\label{e2-1}
	\ee
where $H_0[R]$ is a parametric Hamiltonian with the same properties
as $H[R]$, $\epsilon[R]$ is a real parameter, and $h[R]$ is a Hermitian
operator. If the eigenvalue equation for $H_0[R]$ is exactly solvable, 
then one may attempt to obtain the eigenvalues and eigenvectors of 
$H[R]$ as power series in $\epsilon[R]$,
	\bea
	E_n[R]&=&\sum_{\ell=0}^\infty  E_n^{(\ell)}[R]\:\epsilon[R]^\ell\;,
	\label{e2-2}\\
	|n;R\kt&=&\sum_{\ell=0}^\infty |n;R\kt_\ell\: \epsilon[R]^\ell \;,
	\label{e2-3}
	\eea
whose coefficients $E_n^{(\ell)}[R]$ and $|n;R\kt_\ell$ are expressed in 
terms of the eigenvalues and eigenvectors of $H_0[R]$.

In the following calculations we shall suppress the $R$-dependence of the
relevant quantities for brevity, i.e., we shall use the notation:
	\bea
	&&H=H[R]\;,~~~H_0=H_0[R]\;,~~~ \epsilon=\epsilon[R]\;,~~~~
	h=h[R]\;,\nn\\
	&&E_n=E_n[R]\;,~~~|n\kt=|n;R\kt\;,~~~
	E_n^{(\ell)}=E_n^{(\ell)}[R]\;,~~{\rm and}~~|n\kt_\ell=
	|n;R\kt_\ell\;.\nn
	\eea

Substituting Eqs.~(\ref{e2-2}) and (\ref{e2-3}) in Eq.~(\ref{ei-va-eq})
and performing the necessary calculations, we obtain an equation of the 
form
	\be
	\sum_{\ell=0}^\infty |\xi\kt_\ell \;\epsilon^\ell =0\;,
	\label{e2-4}
	\ee
where
	\bea
	|\xi\kt_0&=&(H_0-E_n^{(0)})|n\kt_0\;,~~~{\rm and}
	\label{e2-5}\\
	|\xi\kt_\ell&=&H_0|n\kt_\ell+h|n\kt_{\ell-1}-\sum_{k=0}^\ell
	E_n^{(k)}|n\kt_{\ell-k}\;,~~~{\rm for}~~~\ell\geq 1\;.
	\label{e2-6}
	\eea
The basic idea of the perturbation theory is to construct a solution of 
Eq.~(\ref{e2-4}) by requiring
	\be
	|\xi\kt_\ell=0\;,~~~{\rm for~all}~~~\ell=0,1,2,\cdots.
	\label{ansatz}
	\ee
For $\ell=0$, this implies that $E_n^{(0)}$ and $|n\kt_0$ are the
eigenvalues and eigenvectors of $H_0$. Therefore, according to the
hypothesis we can calculate them exactly. We shall assume without loss of
generality that $|n\kt_0$ form a complete orthonormal set of basis vectors 
of the Hilbert space and express $|n\kt_\ell$ in this basis. This leads to
	\be
	|n\kt_\ell=\sum_m C_{mn}^\ell |m\kt_0\;,
	\label{e2-7}
	\ee
where $C_{mn}^\ell=C_{mn}^\ell[R]$ are complex coefficients depending
on the parameters $R$. Clearly, $C_{mn}^\ell=~_0\!\br m|n\kt_\ell$. In
particular,
	\be
	C_{mn}^0=\delta_{mn}\;.
	\label{e2-8}
	\ee

Now let us substitute Eq.~(\ref{e2-7}) in Eq.~(\ref{e2-6}) and use 
Eq.~(\ref{e2-8}) and the identity
	\be
	h=\sum_{r,s}~_0\!\br r|h|s\kt_0\: |r\kt_0~_0\!\br s|\;,
	\label{e2-9'}
	\ee
to simplify the resulting expression. This yields 
	\bea
	|\xi\kt_\ell&=&\sum_m d_m^\ell|m\kt_0\;,~~~{\rm where}~~
	\ell\geq 1\;,
	\label{e2-9}\\
	d_m^1&=&-\delta_{mn}E_n^{(1)}+(E_m^{(0)}-E_n^{(0)})
	C^1_{mn}+~_0\!\br m|h|n\kt_0\;,~~~~{\rm and}
	\label{e2-10}\\
	d_m^\ell&=&-\delta_{mn}E_n^{(\ell)}+(E_m^{(0)}-E_n^{(0)})
	C^\ell_{mn}+\sum_r C^{\ell-1}_{rn}~_0\!\br m|h|r\kt_0-
	\sum_{k=1}^{\ell-1}E_n^{(k)}C_{mn}^{(\ell-k)},~
	{\rm for}~\ell\geq 2.
	\label{e2-11}
	\eea

Next we enforce Eq.~(\ref{ansatz}). In view of Eq.~(\ref{e2-9}) and
linear independence of the basis vectors $|m\kt_0$, Eq.~(\ref{ansatz})
implies $d_m^\ell=0$ for all $m$ and $\ell\geq 1$. For $\ell=1$, this
leads to
	\bea
	E_n^1&=& ~_0\br n|h|n\kt_0\;,~~~{\rm and}
	\label{e2-12}\\
	C^1_{mn}&=&\frac{~_0\br m|h|n\kt_0}{E_n^{(0)}-E_m^{(0)}}
	~~~{\rm for}~~~m\neq n.
	\label{e2-13}
	\eea
Eqs.~(\ref{e2-12}) and (\ref{e2-13}) are obtained by setting $m=n$ 
and $m\neq n$ in $d_m^1=0$, respectively. Similarly, $d_m^\ell=0$ for 
$\ell\geq 2$ give rise to
	\bea
	E_n^{(\ell)}&=&\sum_r C_{rn}^{\ell-1}~_0\br n|h|r\kt_0
	-\sum_{k=1}^{\ell-1}E_n^k C_{nn}^{\ell-k},
	 ~~~{\rm for}~~\ell\geq 2
	\label{e2-14}\\
	C^\ell_{mn}&=&(E_n^{(0)}-E_m^{(0)})^{-1}\left(
	\sum_r C_{rn}^{\ell-1}~_0\br m|h|r\kt_0-
	\sum_{k=1}^{\ell-1}E_n^{(k)}C_{mn}^{\ell-k}\right)
	,~~{\rm for}~m\neq n,~\ell\geq 2.
	\label{e2-15}
	\eea
We can use Eqs.~(\ref{e2-12}) and (\ref{e2-13}), to write 
Eqs.~(\ref{e2-14}) and (\ref{e2-15}) in the form
	\bea
	E_n^{(2)}&=&\sum_{r\neq n}(E_r^{(0)}-E_n^{(0)})
	C^1_{nr}C^1_{rn}\;,
	\label{e2-16}\\
	E_n^{(\ell)}&=& \sum_{r\neq n}(E_r^{(0)}-E_n^{(0)})
	C^1_{nr}C^{\ell-1}_{rn}-\sum_{k=2}^{\ell-1}E_n^{(k)} 
	C_{nn}^{\ell-k},~~~{\rm for}~~\ell\geq 3\;,
	\label{e2-17}\\
	C^2_{mn}&=&\sum_{r\neq m}
	\left(\frac{E^{(0)}_r-E^{(0)}_m}{E_n^{(0)}-E_m^{(0)}}
	\right)C^1_{mr} C^1_{rn}+\left(
	\frac{E^{(1)}_m-E^{(1)}_n}{E_n^{(0)}-E_m^{(0)}}
	\right)C^1_{mn}~~~{\rm for}~~m\neq n,
	\label{e2-18}\\
	C^\ell_{mn}&=&\sum_{r\neq m}
	\left(\frac{E^{(0)}_r-E^{(0)}_m}{E_n^{(0)}-E_m^{(0)}}
	\right)C^1_{mr} C^{\ell-1}_{rn}+
	\left(\frac{E^{(1)}_m-E^{(1)}_n}{E_n^{(0)}-E_m^{(0)}}
	\right)C^{\ell-1}_{mn}-	\sum_{k=2}^{\ell-1}
	\frac{E_n^{(k)}C_{mn}^{\ell-k}}{E_n^{(0)}-E_m^{(0)}}\nn\\
	&&\hspace{7cm}
	~~{\rm for}~~m\neq n,\,\ell\geq 3\;.
	\label{e2-19}
	\eea
These equations yield $E_n^{(\ell)}$ and $C_{mn}^\ell$ with $m\neq n$
in terms of  $E_r^{(k)}$, $E_r^{(k)}$, $C_{rs}^k$, and $C_{rr}^k$ 
where $k<\ell$.  One can iterate them to express $E_n^{(\ell)}$ and 
$C_{mn}^\ell$ with $m\neq n$ in terms of  $E_r^{(0)}$, $E_r^{(1)}$, 
$C_{rs}^1$, and $C_{rr}^k$. They do not, however, restrict 
$C_{nn}^\ell$. This means that $C_{nn}^\ell$ are not fixed by the 
eigenvalue equation. This is due to the fact that the eigenvalue equation 
(\ref{ei-va-eq}) determines the eigenvectors up to an arbitrary 
multiplicative factor. We can restrict the choice of $C_{nn}^\ell$ by 
imposing the normalization condition on $|n\kt$. Substituting 
Eqs.~(\ref{e2-3}) and (\ref{e2-7}) in $\br m|n\kt=\delta_{mn}$ and 
making use of the orthonormality of  $|n\kt_0$, we find 
	\bea
	&&\sum_{j=1}^\infty d_j \epsilon^j=0\;,~~~{\rm where}
	\label{e2-20}\\	
	&&d_j:=\sum_{\ell=0}^j\sum_r 	C^\ell_{rn}C^{j-\ell*}_{rm}\;.
	\label{e2-21}
	\eea
Again we seek a solution of Eq.~(\ref{e2-20}) of the form $d_j=0$
for all $j=1,2,\cdots$. This leads to
	\bea
	C^{1*}_{nm}+C^{1}_{mn}&=&0\;,
	\label{e2-21.1}\\
	C^{j*}_{nm}+C^{j}_{mn}&=&
	-\sum_{\ell=1}^{j-1}\sum_r  C^\ell_{rn}C^{j-\ell*}_{rm}\;,~~~
	{\rm for}~~~j\geq 2\;.
	\label{e2-21.2}
	\eea
One can show that for $m\neq n$, Eqs.~(\ref{e2-21.1}) and 
(\ref{e2-21.2}) are trivially satisfied. But for $m=n$, they determine the
real part of  $C^\ell_{nn}$ according to
	\bea
	{\rm Re}(C^1_{nn})&=&0\;,~~~{\rm and}
	\label{e2-22}\\
	{\rm Re}(C^j_{nn})&=&-\frac{1}{2}\sum_{\ell=1}^{j-1}
	\sum_r C^\ell_{rn}C^{j-\ell^*}_{rn}\;,~~~{\rm for}~~
	j\geq 2,
	\label{e2-23}
	\eea	
where ${\rm Re}$ means the `real part of'. The imaginary part of 
$C^\ell_{nn}$ is still arbitrary. This is because the normalization 
condition determines the eigenvectors up to an arbitrary phase factor. 
This phase factor can, in principle, depend on the perturbation parameter
$\epsilon$ and consequently show up in all orders of perturbation. The 
common practice is to set the imaginary part of $C^\ell_{nn}$ equal to 
zero, \cite{dbohm-qm}. This corresponds to making a particular choice for
the phase of the eigenvectors. 
%We shall refrain from making such a choice 
%and retain $C^\ell_{nn}$ in our calculation of the Berry's connection 
%one-form. 

\section{Perturbative Calculation of Berry's Connection One-form}

Having obtained the perturbation series for the eigenvectors $|n\kt$ of
the Hamiltonian $H$, we are in a position to compute the Berry's 
connection one-form $A_n$. In fact, we shall instead compute 
${\cal A}_n$ of Eq.~(\ref{e1-5}). $A_n$ can be easily obtained from
${\cal A}_n$ by changing the time-derivatives to the exterior derivatives.

We shall first substitute Eq.~(\ref{e2-7}) into Eq.~(\ref{e2-3}). This
yields
	\be
	|n\kt=\sum_{\ell=0}^\infty\sum_m C^\ell_{mn}|m\kt_0\epsilon^\ell\;.
	\label{e3-1}
	\ee
Next, we differentiate both sides of Eq.~(\ref{e3-1}) and take the inner 
product of the resulting expression with $|n\kt$. Then using 
Eq.~(\ref{e2-7}), the identity
	\[ ~_k\!\br n|\frac{d}{dt}|m\kt_0=\sum_r~_k\!\br n|r\kt_0
	~_0\!\br r|\frac{d}{dt}|m\kt_0\;,\]
and doing the necessary algebra, we find
	\be
	{\cal A}_n=i\br n|\frac{d}{dt}|n\kt=i\sum_{\ell,k=0}^\infty
	\sum_m\left[ (C^{k*}_{mn}\dot C^\ell_{mn}+\sum_r 
	C^{k*}_{rn}C^\ell_{mn}~_0\!\br r|\frac{d}{dt}|m\kt_0)
	\epsilon^{\ell+k}+\ell C^{k*}_{mn}C^\ell_{mn}\dot\epsilon
	\epsilon^{\ell+k-1}\right],
	\label{e3-2}
	\ee
where a dot denotes a time-derivative. 

Making the change of dummy index: $k\to j:=\ell+k$ we can write 
Eq.~(\ref{e3-2}) in the form
	\bea
	{\cal A}_n&=&\sum_{j=0}^\infty\sum_{\ell=0}^j\sum_m \left[
	i C^{j-\ell*}_{mn}\dot C^\ell_{mn}+
	(C^{j-\ell*}_{mn} C^\ell_{mn} {\cal A}_m^{(0)}) \right]\epsilon^j+
	\nn\\
	&&
	\sum_{j=0}^\infty\sum_{\ell=0}^j\sum_m\sum_{r\neq m}
	C^{j-\ell*}_{rn}C^\ell_{mn}{\cal A}_{rm}^{(0)}\epsilon^j+
	i\sum_{j=1}^\infty\sum_{\ell=1}^j\sum_m
	\ell C^{j-\ell*}_{mn} C^\ell_{mn}\dot\epsilon\epsilon^{j-1},
	\label{e3-4}
	\eea
where ${\cal A}_{rm}^{(0)}:=i~ _0\br r|\frac{d}{dt}|m\kt_0$ and
${\cal A}_m^{(0)}:={\cal A}_{mm}^{(0)}=i~ _0\br m|\frac{d}{dt}
|m\kt_0$.

The last term on the right hand side of (\ref{e3-4}) may be written as
	\[\ell C^{j-\ell*}_{mn} C^\ell_{mn}\dot\epsilon\epsilon^{j-1}=
	\frac{d}{dt}\left[(\frac{\ell}{j})
	C^{j-\ell*}_{mn} C^\ell_{mn}\epsilon^{j}\right]-
	\frac{\ell}{j}\left( \dot C^{j-\ell*}_{mn} C^\ell_{mn}+
	C^{j-\ell*}_{mn} \dot C^\ell_{mn}\right)\epsilon^j\;.\]
Substituting this equation in (\ref{e3-4}), writing the $j=0,1,$ and
$2$ terms in (\ref{e3-4}) separately, and making use of $\dot C^0_{mn}=
\dot\delta_{mn}=0$, we obtain
	\bea
	{\cal A}_n&=& {\cal A}^{(0)}_n+\left[2{\rm Re}(C^1_{nn})
	{\cal A}^{(0)}_n+
	2\sum_{r\neq n}{\rm Re}(C^{1}_{rn}{\cal A}_{nr}^{(0)})
	\right]\epsilon+
	\left[2{\rm Re}(C^2_{nn}) {\cal A}^{(0)}_n+
	\sum_m |C^1_{mn}|^2 {\cal A}^{(0)}_m+\right.\nn\\
	&&\hspace{.3cm}\left.
	\frac{i}{2}\sum_m( C^{1*}_{mn}\dot C^1_{mn}-
	\dot C^{1*}_{mn}C^1_{mn})+
	\sum_{r\neq n} 2{\rm Re}(C^2_{rn}{\cal A}_{nr}^{(0)})
	+\sum_m\sum_{r\neq m}C^{1*}_{rn}C^1_{mn}
	{\cal A}_{rm}^{(0)}\right]\epsilon^2+\nn\\
	&&\hspace{.3cm}
	\sum_{j=3}^\infty\sum_{\ell=0}^j\sum_m\left[
	C^{j-\ell*}_{mn}C^\ell_{mn} {\cal A}^{(0)}_m+
	i(1-\frac{\ell}{j})C^{j-\ell*}_{mn}\dot C^\ell_{mn} -
	\frac{i\ell}{j}\dot C^{j-\ell*}_{mn}C^\ell_{mn}+\right.\nn\\
	&&\hspace{.3cm}\left.
	\sum_{r\neq m}C^{j-\ell*}_{rn}C^\ell_{mn}
	{\cal A}_{rm}^{(0)}\right]\epsilon^j+i\frac{df}{dt}\;,
	\label{e3-5}
	\eea
where 
	\be
	f:=\sum_{j=1}^\infty\sum_{\ell=1}^j\sum_m(\frac{\ell}{j})
	C^{j-\ell*}_{mn}C^\ell_{mn} \epsilon^j\;.
	\label{e3-6}
	\ee
The first term in the first square bracket on the right hand side of  
Eq.~(\ref{e3-5}) vanishes by virtue of Eq.~(\ref{e2-22}). Similarly
using Eq.~(\ref{e2-23}), we can write the first two terms of the second
square bracket in the form:
	\bea
	2{\rm Re}(C^2_{nn}) {\cal A}^{(0)}_n+
	\sum_m |C^1_{mn}|^2 {\cal A}^{(0)}_m
	&=&\sum_m(- |C^1_{mn}|^2  {\cal A}^{(0)}_n+
	|C^1_{mn}|^2 {\cal A}^{(0)}_m)\nn\\
	&=&\sum_{m\neq n}|C^1_{mn}|^2(
	{\cal A}^{(0)}_m-{\cal A}^{(0)}_n)\;.
	\label{e3-7}
	\eea
Next let us observe that in view of Eq.~(\ref{e2-22}),
	\be
	C^{1*}_{nn}\dot C^1_{nn}-C^1_{nn}\dot C^{1*}_{nn}=0.
	\label{e3-7.1}
	\ee
Substituting Eqs.~(\ref{e2-22}), (\ref{e3-7}), and (\ref{e3-7.1}) in 
Eq.~(\ref{e3-5}), we obtain
	\bea
	{\cal A}_n&=& {\cal A}^{(0)}_n+\left[
	2\sum_{r\neq n}{\rm Re}(C^{1}_{rn}{\cal A}_{nr}^{(0)})
	\right]\epsilon+
	\left[\sum_{m\neq n}\left\{ |C^1_{mn}|^2(
	{\cal A}^{(0)}_m-{\cal A}^{(0)}_n)+
	\frac{i}{2}( C^{1*}_{mn}\dot C^1_{mn}-
	\dot C^{1*}_{mn}C^1_{mn})+\right.\right.\nn\\
	&&\hspace{.3cm}\left.\left.
	2{\rm Re}(C^2_{mn}{\cal A}_{nm}^{(0)})\right\}
	+\sum_m\sum_{r\neq m}C^{1*}_{rn}C^1_{mn}
	{\cal A}_{rm}^{(0)}\right]\epsilon^2+
	{\cal O}(\epsilon^3)+i\frac{df}{dt}\;,
	\label{e3-8}
	\eea
where ${\cal O}(\epsilon^3)$ denotes the third and higher order terms in 
$\epsilon$, i.e.,
	\be
	{\cal O}(\epsilon^3):=
	\sum_{j=3}^\infty\sum_{\ell=0}^j\sum_m\left[
	C^{j-\ell*}_{mn}C^\ell_{mn} {\cal A}^{(0)}_m+
	i(1-\frac{\ell}{j})C^{j-\ell*}_{mn}\dot C^\ell_{mn} -
	\frac{i\ell}{j}\dot C^{j-\ell*}_{mn}C^\ell_{mn}+
	\sum_{r\neq m}C^{j-\ell*}_{rn}C^\ell_{mn}
	{\cal A}_{rm}^{(0)}\right]
	\epsilon^j\;.
	\label{third-order}
	\ee
We can rewrite  the terms involving ${\cal A}^{(0)}_m$ in (\ref{third-order})
by separating the $\ell=0$ and $\ell=j$ terms in the sum and using 
Eq.~(\ref{e2-23}) which yields
	\bea
	\sum_{\ell=0}^j\sum_m
	C^{j-\ell*}_{mn}C^\ell_{mn} {\cal A}^{(0)}_m
	&=& 2{\rm Re}(C^j_{nn}){\cal A}^{(0)}_n+
	\sum_{\ell=1}^{j-1}\sum_m
	C^{j-\ell*}_{mn}C^\ell_{mn} {\cal A}^{(0)}_m\,,\nn\\
	&=&\sum_{m\neq n}\sum_{\ell=1}^{j-1}
	C^{j-\ell*}_{mn}C^\ell_{mn} 
	({\cal A}^{(0)}_m-{\cal A}^{(0)}_n)\;.
	\label{e3-8.1}
	\eea
Furthermore, changing the dummy index $\ell$ in the second sum on the
right hand side of (\ref{third-order}) to $k:=j-\ell$, we have
	\be
	\sum_{\ell=0}^j
	i(1-\frac{\ell}{j})C^{j-\ell*}_{mn}\dot C^\ell_{mn}
	=\sum_{k=0}^j\frac{ik}{j}C^{k*}_{mn}\dot C^{j-k}_{mn}\;.
	\label{e3-8.2}
	\ee 
Substituting Eqs.~(\ref{e3-8.1}) and (\ref{e3-8.2}) in (\ref{third-order}),
we find
	\bea
	{\cal O}(\epsilon^3) &=&
	\sum_{j=3}^\infty\sum_{\ell=1}^{j-1}
	\left[\sum_{m\neq n}
	C^{j-\ell*}_{mn}C^\ell_{mn} ({\cal A}^{(0)}_m-
	{\cal A}^{(0)}_n)+\sum_m\left\{
	\frac{i\ell}{j}(C^{j-\ell*}_{mn}\dot C^\ell_{mn} -
	\dot C^{j-\ell*}_{mn}C^\ell_{mn})+\right.\right.\nn\\
	&&\hspace{.3cm}\left.\left.
	\sum_{r\neq m}C^{j-\ell*}_{rn}C^\ell_{mn}
	{\cal A}_{rm}^{(0)}\right\}\right]
	\epsilon^j\;.
	\label{third-order.1}
	\eea

Having obtained the perturbation series for ${\cal A}_n$ we can write
down the  perturbation series for the Berry's connection one-form $A_n$.
Changing the time-derivatives to the exterior derivatives in the
expression (\ref{e3-8}) for ${\cal A}_n$, we find
	\bea
	A_n&=&A^{(0)}_n+
	\left[2\sum_{r\neq n}{\rm Re}(C^{1}_{rn}A_{nr}^{(0)})
	\right]\epsilon+
	\left[\sum_{m\neq n}\left\{ |C^1_{mn}|^2(A^{(0)}_m- A^{(0)}_n)+
	\frac{i}{2}( C^{1*}_{mn} dC^1_{mn}-dC^{1*}_{mn}C^1_{mn})+
	\right.\right.\nn\\
	&&\hspace{.3cm}\left.\left.
	2{\rm Re}(C^2_{mn}A_{nm}^{(0)})\right\}
	+\sum_m\sum_{r\neq m}C^{1*}_{rn}C^1_{mn}
	A_{rm}^{(0)}\right]\epsilon^2+
	O(\epsilon^3)+idf\;,
	\label{e3-9}
	\eea
where 
	\bea
	A^{(0)}_{mn}&:=&i~_0\!\br m|d|n\kt_0\;,~~~~
	A^{(0)}_{n}:=A^{(0)}_{nn}=i~_0\!\br n|d|n\kt_0\;,~~~{\rm and}
	\label{e3-9.1}\\
	O(\epsilon^3)&:=&
	\sum_{j=3}^\infty\sum_{\ell=1}^{j-1}
	\left[\sum_{m\neq n}
	C^{j-\ell*}_{mn}C^\ell_{mn} ( A^{(0)}_m-
	A^{(0)}_n)+\sum_m\left\{
	\frac{i\ell}{j}(C^{j-\ell*}_{mn}\,dC^\ell_{mn} -
	dC^{j-\ell*}_{mn}C^\ell_{mn})+\right.\right.\nn\\
	&&\hspace{.3cm}\left.\left.
	\sum_{r\neq m}C^{j-\ell*}_{rn}C^\ell_{mn}
	A_{rm}^{(0)}\right\}\right]
	\epsilon^j\;.
	\label{third-order-form}
	\eea

In particular, let us consider a case where $A^{(0)}_{mn}=0$ for all $m$ and $n$. Then,
	\be
	A_n=\sum_{m\neq n}\frac{i}{2}( C^{1*}_{mn} dC^1_{mn}-
	dC^{1*}_{mn}C^1_{mn})\epsilon^2+
	O(\epsilon^3)+idf\;.
	\label{e3-9-0}
	\ee
If the unperturbed Hamiltonian $H_0$ is a fixed operator, its eigenvectors
$|n\kt_0$ will not depend on $R$. In this case $A^{(0)}_{mn}=0$ and 
Eq.~(\ref{e3-9-0}) holds. This equation indicates that {\em the 
geometric phase effects due to a time-dependent perturbation are second 
(or higher) order effects in the perturbation parameter.} In fact, this
statement is also valid for the general case where
$A_{mn}^{(0)}\neq 0$. In order to see this, we recall the
hypothesis of the adiabaticity of the evolution \cite{pra97a} which requires 
	\be
	{\cal A}_{mn}:=i\br m|\frac{d}{dt}|n\kt\approx 0
	~~~~{\rm for~all}~~~m\neq n\,. 
	\label{e3-9.1'}
	\ee
We can repeat the above calculation of ${\cal A}_n$ for ${\cal A}_{mn}$
with $m\neq n$ and show that 
	\[{\cal A}_{mn}={\cal A}_{mn}^{(0)}+\mbox{terms of order 
	$\epsilon$ and higher}.\] 
Hence, in order to ensure the validity of the adiabaticity condition 
(\ref{e3-9.1'}), ${\cal A}_{mn}^{(0)}$ with $m\neq n$ must be at least 
of order $\epsilon$. Consequently, the first perturbative correction to Berry's 
connection one-form (\ref{e3-9}) is indeed of order $\epsilon^2$.

\section{Particle in a One-dimensional Infinite Well with Moving
Boundaries}

The Schr\"odinger equation for a particle of mass $M$ in a 
one-dimensional infinite square well with a moving boundary 
is given by
	\bea
	i\hbar\dot\psi(x;t)&=&\left[-\frac{\hbar^2}{2M}\,
	\frac{\partial^2}{\partial x^2}+V(x,t)\right]\psi(x;t)\;,
	\label{e1}\\
	\psi(0;t)&=&\psi(L(t);t)=0\;,
	\label{e2}
	\eea
where $V(x,t)$ is a real interaction potential, $L:[0,\tau]\to\R^+$ is a 
smooth function, $\tau$ is the duration of the evolution of the system,
and $x=0$ and $x=L(t)$ are the positions of the boundaries.

As argued by Pereshogin and Pronin \cite{pe-pr}, who studied the case
of a free particle ($V=0$), the Hilbert space ${\cal H}_t$ of this system at
time $t$ is $L^2([0,L(t)])$. In particular, ${\cal H}_t$ is time-dependent. One way
to handle this situation is to identify ${\cal H}_t$ with a fiber of a vector 
bundle, endow this vector bundle with a connection, and replace the 
ordinary time derivative appearing in the Schr\"odinger equation~(\ref{e1})
by the covariant time-derivative corresponding to this connection. This is 
the approach pursued by Pereshogin and Pronin \cite{pe-pr}. If one makes 
the same choice for the connection as the one made by Pereshogin and 
Pronin \cite{pe-pr}, then one obtains the effective Hamiltonian
	\be
	H_{\rm eff}(t)=\frac{p^2}{2M}+\frac{\dot L(t)}{2L(t)}
	(xp+px)\;,
	\label{e3}
	\ee
which is valid for $V=0$. Pereshogin and Pronin suggest that the dynamics
of such a particle is determined by the Schr\"odinger equation for this 
effective Hamiltonian subject to the same boundary conditions as in 
(\ref{e2}).\footnote{As we shall see below, a consistent treatment of this 
problem leads to an effective Hamiltonian which differs from $H_{\rm eff}$ 
in the sign of the second term on the right hand side of (\ref{e3}).}

The conventional approach to this problem is to determine the dynamics
of the system using the Hamiltonian \cite{do-ri} 
	\bea
	H(t)&=&\frac{p^2}{2M}+\tilde V(x;t)\;,~~~{\rm where}
	\label{e4}\\
	\tilde V(x,t)&=&\left\{\begin{array}{ccc}
	V(x,t)&{\rm for}& x\in[0,L(t)]\\
	\infty&{\rm for}&x\notin [0,L(t)]
	\end{array}\right..
	\label{e5}
	\eea
The Schr\"odinger equation for this Hamiltonian is clearly equivalent to the
original Schr\"odinger equation~(\ref{e1}).

We shall approach the problem of solving the Schr\"odinger equation for 
this system by applying the time-dependent canonical transformation 
\cite{jmp97b,jpa98a},
	\bea
	|\psi(t)\kt&\to& |\psi'(t)\kt:= {\cal U}(t)  |\psi(t)\kt\;,
	\label{trans-psi}\\
	H(t)&\to&H'(t):= {\cal U}(t) H(t) {\cal U}(t)^\dagger +
	i \hbar{\cal U}(t)\dot{\cal U}(t)^\dagger\;,
	\label{trans-H}\\
	x&\to&x':= {\cal U}(t) \: x\: {\cal U}(t)^\dagger\;,~~~~
	p~\to~p':= {\cal U}(t) \: p \:{\cal U}(t)^\dagger\;,
	\label{trans-xp}
	\eea
defined by the unitary operator
	\be
	{\cal U}(t):= e^{\frac{i a(t)}{2\hbar}(xp+px)}\;,
	\label{u}
	\ee
where $a=a(t)$ is a smooth real-valued function of time. This canonical
transformation corresponds to a time-dependent dilatation of space 
\cite{jpa98a}. This is easily seen by substituting (\ref{u}) in 
(\ref{trans-xp}) which yields
	\be
	x\to x'= e^{a(t)}  x\;,~~~{\rm and}~~~
	p\to p'= e^{-a(t)} p\;.
	\label{e10}
	\ee
Furthermore, substituting (\ref{u}) in (\ref{trans-H}) and using 
Eq.~(\ref{e4}), we find
	\be
	H(t)\to H'(t)=\frac{p^{'2}}{2M}+\tilde V'(x,t)
	-\frac{\dot a(t)}{2}(xp+px)\;,
	\label{e11}
	\ee
where
	\be
	\tilde V'(x,t):= \tilde V(x',t)\;.
	\label{e11.1}
	\ee

Next let us choose the dilatation parameter $a(t)$ to be
	\be
	a(t)=\ln(L(t)/L_0)\;,
	\label{e12}
	\ee
where $L_0:=L(0)$. Substituting (\ref{e12}) in (\ref{e10}), we
obtain
	\be
	x'=\left(\frac{L(t)}{L_0}\right)x\;,~~~{\rm and}~~~
	p'=\left(\frac{L_0}{L(t)}\right)p\;.
	\label{e13}
	\ee
In view of Eqs.~(\ref{e5}), (\ref{e11.1}), and (\ref{e13}), the transformed
potential is given by
	\be
	\tilde V'(x,t)=\tilde V(x',t):=\left\{\begin{array}{ccc}
	V(x',t)&{\rm for}& x'\in[0,L(t)]\\
	\infty&{\rm for}&x'\notin [0,L(t)]
	\end{array}\right\}=\left\{\begin{array}{ccc}
	V(\frac{L(t)x}{L_0} ,t)&{\rm for}& x\in[0,L_0]\\
	\infty&{\rm for}&x\notin [0,L_0]
	\end{array}\right..
	\label{e14}
	\ee
This means that the Schr\"odinger equation for the transformed 
Hamiltonian $H'(t)$ is equivalent to the Schr\"odinger equation for the
Hamiltonian
	\be
	H''(t)=\frac{L_0^2p^2}{2ML(t)^2}-\frac{\dot L(t)}{2L(t)}(xp+px)
	+V(\frac{L(t)x}{L_0},t)\;,
	\label{e15}
	\ee
namely
	\be
	i\hbar\dot\psi'(x;t)=
	\left[-\frac{\hbar^2L_0^2}{2ML(t)^2}\frac{\partial^2}{\partial x^2}
	+\frac{i\hbar\dot L(t)}{2L(t)}(x\frac{\partial}{\partial x}
	+\frac{\partial}{\partial x}x)+V(\frac{L(t)x}{L_0},t)\right]\psi'(x;t)\;,
	\label{e16}
	\ee
where $\psi'(x;t)\in L^2([0,L_0])$ and (\ref{e16}) is supposed to be solved
with boundary conditions
	\be
	\psi'(0;t)=\psi'(L_0;t)=0\;.
	\label{e17}
	\ee
The canonical transformation defined by (\ref{u}) and (\ref{e12}), 
therefore, maps the dynamics of the system with a time-dependent 
configuration space, i.e., $[0,L(t)]$, to a system with a constant 
configuration space, i.e., $[0,L_0]$. The idea of transforming the problem
with moving boundaries to an equivalent one with fixed boundaries was 
previously used by Munier {\em et al} \cite{mu-bu-fe-fi}, Razavy 
\cite{razavy}, Greenbereger \cite{greenberger}, and Seba \cite{seba}.

We conclude this section by making a couple of remarks.
	\begin{itemize}
	\item[1.] Performing the canonical transformation (\ref{u}) and 
(\ref{e12}), on the effective Hamiltonian (\ref{e3}) of Pereshogin 
and Pronin \cite{pe-pr}, we obtain the transformed effective Hamiltonian
	\be
	H''_{\rm eff}(t)=\frac{L_0^2p^2}{2ML(t)^2}\;.
	\label{e18}
	\ee
This is the Hamiltonian of a particle with a time-dependent
(effective) mass $\tilde M(t)=ML^2(t)/L_0^2$ which is confined between two
walls positioned at $x=0$ and $x=L_0$. The Schr\"odinger equation for
$H''_{\rm eff}(t)$ can be easily solved, for the adiabatic approximation
yields the exact result \cite{pra97b}. This means that the approach of 
Pereshogin and Pronin \cite{pe-pr} leads to a Hamiltonian that is
canonically equivalent to that of a free particle with a time-dependent mass.
The eigenvalue problem for $H''_{\rm eff}(t)$ is also solved exactly and 
there is no need to appeal to perturbation theory.
	\item[2.] In the Schr\"odinger equation (\ref{e16}) for the transformed
Hamiltonian (\ref{e15}), if one combines the term $i\hbar {\cal U}(t)
\dot{\cal U}(t)^\dagger=-\dot L(t)(xp+px)/(2L(t))$ with the time derivative, 
one obtains the `covariant time derivative' 
	\be
	\nabla_t':=\frac{\partial }{\partial t}-\frac{\dot L}{2 L}\left(
	x\frac{\partial}{\partial x}+\frac{\partial}{\partial x}\,x\right)\;.
	\label{cov-der}
	\ee
The covariant time derivative $\nabla_t$ of Eq.~(9) of Pereshogin and Pronin 
\cite{pe-pr} differs from (\ref{cov-der}) by a minus sign in the second term on 
the right hand side of (\ref{cov-der}).  Indeed as pointed out by one of the 
referees, a consistent treatment of the problem based on the method of 
Pereshogin and Pronin \cite{pe-pr} shows that in fact (\ref{cov-der}) is the 
correct expression for the covariant time derivative. In order to see this, one 
must reconsider the definition of the operator $\hat P:L^2([0,L(t_1)])\to
L^2([0,L(t_2)])$ of Pereshogin and Pronin \cite{pe-pr} which is used to define
$\nabla_t$. Pereshogin and Pronin determine $\hat P$ by requiring that its 
effect on the wave function (in the coordinate representation) is that of
a dilatation. It is not difficult to see that $\hat P={\cal U}(t)$ where ${\cal U}(t)$
is given by Eq.~(\ref{u}) with $\alpha(t)=\ln[L(t_2)/L(t_1)]$. Note that for 
$\psi(x,t_1)\in L^2([0,L(t_1)])$, $\hat P\psi_n(x,t_1)=\psi_n(x',t_1)$ where 
$x'=[L(t_2)/L(t_1)]x\in[0,L(t_2)]$. Hence, $\hat P\psi_n(x,t_1)\in 
L^2([0,L(t_2)])$, as required.\footnote{Note that here $\hat P$ is an active
transformation: $|\psi\kt\to \hat P|\psi\kt$. It can also be viewed as a 
passive transformation $\br x|\to\br x|\hat P=:\br x'|$.}
Setting $t_1=t+\delta t$ and $t_2=t$, one obtains the infinitesimal form 
of  $\hat P:L^2([0,L(t+\delta t)])\to L^2([0,L(t)])$ which is given by
	\be
	\hat P=1-\delta t\:\frac{\dot L}{2L}\left(x\frac{\partial}{\partial x}
	+\frac{\partial}{\partial x}\,x\right)\;.
	\label{p=}
	\ee
Pereshogin and Pronin's expression for $\hat P$ differs from (\ref{p=}) in the 
sign of the second term on the right hand side of (\ref{p=}). If one chooses the 
opposite sign, as Pereshogin and Pronin do, then $\hat P\psi_n(x,t_1)\notin 
L^2([0,L(t_2)])$, and the construction is inconsistent. If one uses the 
expression (\ref{p=}) for $\hat P$ in the Pereshogin and Pronin's analysis, one 
obtains the covariant time derivative (\ref{cov-der}) and the effective 
Hamiltonian
	\be
	H'_{\rm eff}=\frac{p^2}{2M}-\frac{\dot L(t)}{2L(t)}\,
	(xp+px)\;.
	\label{e3-new}
	\ee
Note that again the relevant Hilbert space is $L^2[([0,L(t)])$.
\end{itemize}

\section{Adiabatic Geometric Phase Due to Moving Boundaries}

\subsection{Adiabatic geometric phase for the Hamiltonian~(\ref{e4})}

The Hamiltonian (\ref{e4}) is a special case of a parametric Hamiltonian
of the form
	\bea
	H[R]&=&\frac{p^2}{2M}+\tilde V(x;R]\;,~~~{\rm where}
	\label{e4.1}\\
	\tilde V(x;R]&=&\left\{\begin{array}{ccc}
	V(x,R]&{\rm for}& x\in[0,L]\\
	\infty&{\rm for}&x\notin [0,L]
	\end{array}\right.,
	\label{e5.1}
	\eea
$R=(L,R^1,\cdots,R^d)$ are real parameters, and
$R^1,\cdots,R^d$  may be viewed as a set of coupling constants 
occurring in the expression for $V$.

For the case of a free particle $V=0$, and one can easily solve the 
eigenvalue equation for this Hamiltonian. The eigenvalues and 
eigenfunctions are given by
	\bea
	E_n&=&\frac{\hbar^2\pi^2 n^2}{2ML^2}\;,~~~{\rm and}
	\label{e6}\\
	\phi_n&=&\sqrt{\frac{2}{L}}
	\sin\left(\frac{\pi n x}{L}\right) [\theta(x)-\theta(x-L)]\;,
	\label{e7}
	\eea
respectively. Since the eigenfunctions are real, one expects Berry's 
connection one-form 
	\be
	A_n:=i\br\phi_n|d|\phi_n\kt\;,
	\label{e8.0}
	\ee
to vanish identically \cite{pla97b}. This is in fact the case for any real 
potential $V(x;t)$, because for a real potential the eigenfunctions 
$\phi_n$ may be chosen to be real. More specifically, one has
	\be
	\phi_n(x;R]=f_n(x,R]  [\theta(x)-\theta(x-L)]\;,
	\label{e8.1}
	\ee
where $f_n$ are real-valued functions depending on $R$ and
vanishing at $x=0$ and $x=L$. A simple calculation shows that
	\bea
	A_n&=&i\int_{-\infty}^\infty dx\left(	f_n 
	\sum_{a=0}^n\frac{\partial f_n}{\partial R^a}dR^a\:
	[\theta(x)-\theta(L-x)]^2+(f_n)^2[\theta(x)-\theta(L-x)]
	\delta(x-L) dL\right)\;,\nn\\
	&=&\frac{i}{2}\int_{0}^L dx\left( 
	\sum_{a=1}^n \frac{\partial (f_n)^2}{\partial R^a}dR^a+
	\frac{\partial (f_n)^2}{\partial L}dL
	+(f_n)^2[\theta(x)-\theta(L-x)]\delta(x-L) dL\right)\;,\nn\\
	&=&\frac{i}{2}\left[
	\left(\sum_{a=1}^n dR^a\frac{\partial}{\partial R^a}+
	dL \frac{\partial}{\partial L}\right)
	\left(\int_0^L f_n^2 dx\right)-dL\,\left.(f_n)^2\right|_{x=L}+
	\right.\nn\\
	&&\hspace{3mm}\left.
	dL[\theta(L)-\theta(0)]\left.(f_n)^2\right|_{x=L}\right]\;,
	\label{e8.2}\\
	&=&0\;.
	\label{e8.3}
	\eea
The integral on the right hand side of (\ref{e8.2}) is the norm of $\phi_n$ 
which is supposed to be one. Therefore, its derivatives vanish. The last
two terms vanish, because $f_n|_{x=L}=0$.

Eq.~(\ref{e8.3}) shows that the problem of the adiabatic geometric phase 
for the Hamiltonian~(\ref{e4}) is trivial. This means that the cyclic adiabatic
geometric phase angles $\gamma_n(T)$ vanish, and the noncyclic adiabatic
geometric phases $\Phi_n(t)$ only depend on the end points of the path traced
by the parameters in the parameter space. 
	
\subsection{Adiabatic geometric phase for the canonically transformed 
Hamiltonian~(\ref{e15}) with $V=0$}

The Hamiltonian~(\ref{e15}) with $V=0$ is obtained from the parametric Hamiltonian
	\be
	H''[L,R]=\frac{L_0^2p^2}{2M L^2}-\frac{R}{2ML^2}(xp+px)\;,
	\label{e6-6}
	\ee
by setting $L=L(t)$ and $R=R(t)=ML(t)\dot L(t)$. The Hilbert space of the system
is $L^2([0,L_0])$.

We shall write $H''$ in the form
	\be
	H''[L,R]=H_0[L]+\epsilon[R] h[L]\;,
	\label{e6-7}
	\ee
where
	\be
	H_0[L]:=\frac{L_0^2p^2}{2M L^2}\;,~~~
	\epsilon[R]:=R\;,~~{\rm and}~~ h[L]:=-\frac{xp+px}{2ML^2}\;.
	\label{e6-8}
	\ee

The eigenvalues and eigenfunctions of $H_0$ are given by
	\be
	E_n^{(0)}=\frac{\hbar^2\pi^2 n^2}{2M L^2}\;,~~
	{\rm and}~~\psi_n^{(0)}(x)=\br x|n\kt_0=\sqrt{\frac{2}{L_0}}\,
	\sin\left(\frac{\pi n x}{L_0}\right)\;,
	\label{e6-9}
	\ee
respectively. Because $\psi_n^{(0)}(x)$ do not depend on $R$ or $L$,
$A_{mn}^{(0)}=0$ and the Berry connection one-form is given by
Eq.~(\ref{e3-9-0}). Using Eqs.~(\ref{e2-12}), (\ref{e2-13}) and 
(\ref{e6-9}) and performing the necessary algebra, we find
	\be
	E^{(1)}_n=0\;,~~{\rm and}~~
	C^1_{mn}=\frac{4i(-1)^{m+n}mn}{\hbar\pi^2(m^2-n^2)^2}\;,~~
	{\rm for}~~m\neq n\;.
	\label{e6-10}
	\ee
Note that $C^1_{mn}$ do not involve $L$. Furthermore, in view of 
Eqs.~(\ref{e2-16}) -- (\ref{e2-19}), $E_n^{\ell}$ will all be either zero or 
proportional to $L^{-2}$. Therefore, their ratios will also be independent 
of $L$. This in turn implies that all $C^\ell_{mn}$  should 
be independent of $L$. Hence $dC^\ell_{mn}=0$ for all $\ell$. 
In view of Eq.~(\ref{e3-9-0})  and $A_{mn}^{(0)}=0$, this is 
sufficient to conclude that $A_n$ is an exact one-form and the geometric
phase is trivial.

\subsection{Adiabatic geometric phase for the effective 
Hamiltonians~(\ref{e3}) and (\ref{e3-new})}

It is not difficult to see that the effective Hamiltonians (\ref{e3}) and 
(\ref{e3-new}) are special cases of a parametric Hamiltonian of the form
	\be
	H_{\rm eff}[R]=\frac{p^2}{2M}+\frac{R}{2}\,(xp+px)\;,
	\label{e3-param}
	\ee
where $R\in\R$ is a real parameter. The effective Hamiltonians (\ref{e3}) 
and (\ref{e3-new}) are obtained from (\ref{e3-param}) by requiring $R$ to 
change in time according to  
	\be
	R(t)=\pm\frac{\dot L(t)}{L(t)}\;. 
	\label{e6-2'}
	\ee
where plus sign corresponds to the effective Hamiltonian (\ref{e3}) and the
minus sign to the effective Hamiltonian (\ref{e3-new}). In the following we
shall only treat the case of the effective Hamiltonian (\ref{e3}). The
analogous results are obtained for the effective Hamiltonian (\ref{e3-new})
by changing the sign of $R$ in the relevant equations.

It is well-known, at least for the cases where the Hilbert space is  $L^2(\R)$, 
that a parametric Hamiltonian which has $\R$ as its parameter space cannot
lead to a nontrivial geometric phase. This is simply because in this case 
Berry's connection one-form  $A_n$ depends on a single variable $R\in\R$ 
and can be written as $dF(R)$ where $F(R)=\int A_n(R)dR$. This implies 
that both cyclic and noncyclic adiabatic geometric phases are trivial. The 
same conclusion can also be reached for the cases that the Hilbert space is 
$L^2({\cal M})$ where ${\cal M}$ is a fixed configuration space.  

The configuration space of the effective Hamiltonian (\ref{e3}) is the 
interval $[0,L]$ which is variable. We can treat this case by identifying the
Hilbert space $L^2([0,L])$ with $L^2_\mu(\R)$ of Eq.~(\ref{L2-mu})
where $\mu=\theta(x)-\theta(x-L)$. In this way, it is clear that the 
expression for the Berry connection one-form involves two parameters,
namely $R$, which enters through the dependence of the 
eigenfunctions of $H_{\rm eff}[R]$ on $R$, and $L$ which 
enters through the dependence of the measure $\mu$ on $L$. Therefore,
the above argument does not apply to $H_{\rm eff}[R]$. 

Following Pereshogin and Pronin \cite{pe-pr}, we compute the 
eigenfunctions of $H_{\rm eff}[R]$ using perturbation theory. 
We shall write
	\be
	H_{\rm eff}=H_0+\epsilon h\;,
	\label{e6-1}
	\ee
where
	\be
	H_0=\frac{p^2}{2M}\;,~~~\epsilon=ML^2 R\;,~~{\rm and}~~
	h=\frac{1}{2ML^2}(xp+px)\;.
	\label{e6-2}
	\ee
The eigenvalues and eigenfunctions of $H_0$ are given by
	\be
	E_n^{(0)}=\frac{\hbar^2\pi^2 n^2}{2ML^2}\;,~~{\rm and}~~
	\psi_n^{(0)}(x)=\br x|n\kt_0=
	\sqrt{\frac{2}{L}}\,\sin\left(\frac{\pi n x}{L}\right)\;.
	\label{e6-3}
	\ee
Because the eigenfunctions $\psi_n^{(0)}(x)$ of $H_0$ are real, and
$\psi_n^{(0)}(L)=0$, we have $A_n^{(0)}=0$. Furthermore, we can
use Eqs.~(\ref{e6-3}), (\ref{e2-12}), (\ref{e2-13}) and ({\ref{e6-2}), to calculate
	\bea
	A_{mn}^{(0)}&=&i_0\br m|\frac{d}{dt}|n\kt_0=
	\frac{2imn(-1)^{m+n}dL}{(m^2-n^2)L}\;,
	~~{\rm for}~~~m\neq n\;,
	\label{e6-20}\\
	E^{(1)}_n&=&0\;,~~~{\rm and}~~~C_{mn}^1=
	\frac{4i(-1)^{m+n+1}mn}{\hbar\pi^2(m^2-n^2)^2} 
	\;,~~~{\rm for}~~m\neq n\;.
	\label{e6-4}
	\eea
Again one can show that $E^{(\ell)}_n$ are all proportional to $L^{-2}$, and
$C^{\ell}_{mn}$ are independent of the parameters. Hence, in view of
Eq.~(\ref{e3-9}), we have
	\bea
	A_n&=&	\left[2\sum_{m\neq n}{\rm Re}(C^{1}_{mn}
	A_{nm}^{(0)})\right]\epsilon+
%	\left[\sum_{m\neq n}2{\rm Re}(C^2_{mn}
%	A_{nm}^{(0)})+\sum_m\sum_{r\neq m}C^{1*}_{rn}C^1_{mn}
%	A_{rm}^{(0)}\right]\epsilon^2+
%	\tilde{\cal O}(\epsilon^3)+idf\;,
%	\label{e3-9'}\\
	\cdots\;,\nn\\
	&=&\left[
	\left(\sum_{m\neq n}\frac{m^2n^2}{(m^2-n^2)^3}\right)
	\left(\frac{-16 dL}{\pi^2\hbar L}\right)\right]\epsilon+\cdots
	\;,\nn\\
	&=&\left(\sum_{m\neq n}\frac{m^2n^2}{(m^2-n^2)^3}\right)
	\left(\frac{-16 M R L dL}{\pi^2\hbar}\right)+\cdots
	\;,\nn\\
	&=&\left(\sum_{m\neq n}\frac{m^2n^2}{(m^2-n^2)^3}\right)
	\left(\frac{-16 M\dot LdL}{\pi^2\hbar}\right)+\cdots\;,
	\label{e3-9''}
	\eea
where `$\cdots$' stands for the terms which are either exact forms or
of higher order in $\epsilon$. Eq.~(\ref{e3-9''}) concides with the result
of Pereshogin and Pronin \cite{pe-pr}. Note that $\Gamma^n_{tn}$ of
Pereshogin and Pronin \cite{pe-pr} is equal to $-{\cal A}_n=-A_n/dt$.

It is worth mentioning that in view of Eqs.~(\ref{e6-2}), (\ref{e6-2'}), and
(\ref{e6-20}), both ${\cal A}_{mn}^{(0)}$ with $m\neq n$ and $\epsilon$  are 
proportional to $\dot L$. This shows that the choice made for the perturbation
parameter $\epsilon$ is consistent with the adiabaticity of the evolution. In 
other words, the perturbation theory is valid for an adiabatic evolution of the
system where $\dot L$ is very small. 

\section{Conclusion}

In this article we have addressed two problems. First we presented a
systematic perturbative calculation of the Berry's connection one-form and 
showed that the Berry's phase due to a time-dependent perturbation is a 
second order effect in the perturbation parameter. Next, we studied the 
quantum dynamics of a particle confined between moving walls and 
reconsidered the problem of the adiabatic geometric phase for this system. 

We showed that using the conventional approach based on the Hamiltonian 
(\ref{e4}) this system does not involve any nontrivial adiabatic geometric 
phases. For the case of a free particle where $V=0$, transforming this 
system into a canonically equivalent one with fixed boundaries does not lead 
to nontrivial adiabatic geometric phases either. However, if one postulates a 
new effective Hamiltonian for the system, then in principle nontrivial geometric
phases may arise even for the case of a free particle. For example, if one
uses the effective Hamiltonian (\ref{e3}) or (\ref{e3-new}), one obtains
a nontrivial adiabatic geometric phase. The effective Hamiltonian (\ref{e3})
turns out to be  canonically equivalent to that of a free particle of variable 
mass which is confined to an infinite square well with fixed boundaries.
The latter system can be solved exactly. In particular, one can show that
it does not involve nontrivial adiabatic geometric phases. The occurrence
of nontrivial adiabatic geometric phases for the effective Hamiltonian
(\ref{e3}) has, therefore, its origin in the time-dependent canonical 
transformation relating the two systems \cite{ga-go-ro-th}. 

We argued that a consistent treatment of the dynamics of a particle confined 
between a fixed and a moving boundary using the method of Pereshogin and 
Pronin \cite{pe-pr} leads to an effective Hamiltonian which differs slightly
from that obtained by Pereshogin and Pronin. The occurrence of nontrivial
geometric phases for this effective Hamiltonian is a clear indication of the fact
that the approach of Pereshogin and Pronin is not equivalent to the 
conventional approach. Since both approaches aim to describe the dynamics
of the same physical system, an experimental investigation of their predictions
can easily determine their validity.

Finally, we wish to remark that the dynamics of a massless particle confined
between moving boundaries can also be treated by transforming the problem
to an equivalent one with fixed boundaries \cite{razavy}. The phenomenon of 
the geometric phase for such a particle has not been addressed in a satisfactory
manner\footnote{See however \cite{le-le}.}, though the geometric phases in
optical systems have been thoroughly investigated. See for example the review
\cite{optics}. The relativistic analog of the adiabatic geometric phase has
been discussed in Ref.~\cite{jpa98b}. The results of \cite{jpa98b} may also be
used to treat the case of massless particles satisfying the wave equation.


\begin{thebibliography}{99}
\bibitem{pe-pr} P.~Pereshogin and P.~Pronin, Phys.~Lett.~A {\bf 156},
12 (1991).
\bibitem{berry1984} M.~V.~Berry, Proc.\ Roy.\ Soc.\ London 
{\bf A 392}, 45 (1984).
\bibitem{do-ri} S.~W.~Doescher and M.~H.~Rice, Am.\ J.\ Phys.\
{\bf 37}, 1246 (1969). 
\bibitem{mu-bu-fe-fi} A.~Munier, J.\ R.\ Burgan, M.\ Feix, and
E.\ Fijalkow, J.~Math.~Phys.\ {\bf 22}, 1219 (1981).
\bibitem{be-kl} M.~V.~Berry and G.\ Klein,  J.~Phys.\ A: Math.\ Gen.\
{\bf 17}, 1805 (1984).
\bibitem{greenberger} D.~M.~Greenberger, Physica B {\bf 151}, 
374 (1988).
\bibitem{pinder} D.~N.~Pinder, Am.\ J.\ Phys.\ {\bf 58}, 54 (1990). 
\bibitem{seba} P.~Seba, Phys.~Rev.~A {\bf 41}, 2306 (1990).
\bibitem{ma-de} A.~J.~Makowski and S.~T.~Dembinski, Phys.~Lett.~A
{\bf 154}, 217 (1991);\\
A.~J.~Makowski and P.~Peplowski, Phys.~Lett.~A
{\bf 163}, 142 (1992);\\
A.~J.~Makowski, J.~Phys.\ A: Math.\ Gen.\ {\bf 25},
3419 (1992).
\bibitem{de-po} A.~Devoto and B.~Pomorisac, J.~Phys.\ A: Math.\ Gen.\
{\bf 25}, 241 (1992).
\bibitem{do-kl-ni} V.~V.~Dodonov, A.~B.~Klimov, and D.~E.~Nikonov,
J.~Math.~Phys.\ {\bf 34}, 3391 (1993).
\bibitem{le-le} J.-M.~Levy-Leblond, Phys.~Lett.~A {\bf 125}, 441
(1987).
\bibitem{aa} Y.~Aharonov and J.~Anandan, Phys.~Rev.~Lett.\ {\bf 58},
1593 (1987)\\
J.~Anandan and Y.~Aharonov, Phys.~Rev.~{\bf D38}, 1863 (1988).
\bibitem{jmp97b} A.~Mostafazadeh, J.~Math.~Phys.\ {\bf 38}, 
3489 (1997)
\bibitem{jpa98a} A.~Mostafazadeh, J.~Phys.\ A: Math.\ Gen.\ {\bf 31}, 
6495 (1998)
\bibitem{adiabatic} M.~Born and V.~Fock, Zeit.~F.~Phys.\ {\bf 51}, 
165 (1928);\\
T.~Kato, J.\ Phys.\ Soc.\ Jpn.\ {\bf 5}, 435 (1950).
\bibitem{pra97a} A.~Mostafazadeh, Phys.~Rev.~A {\bf 55}, 1653 (1997).
\bibitem{p31}  A.~Mostafazadeh, ``Noncyclic Geometric Phase and Its
Non-Abelian Generalization,'' J.~Phys.~A: Math.\ Gen., to appear.
\bibitem{po-sj}  G.~G.~de Polavieja and E.~Sj\"oqvist, Am.\ J.\ Phys.\ 
{\bf 66}, 431 (1998).
\bibitem{dbohm-qm} D.~Bohm, {\em Quantum Theory} (Dover,
New York, 1989).
\bibitem{abohm-qm} A.~Bohm, {\em Quantum Mechanics:
Foundations and Applications,} third edition (Springer-Verlag, 
Berlin, 1993).
\bibitem{razavy} M.~Razavy, Hadronic J.\ {\bf 8}, 153 (1985);\\
M.~Razavy, Phys.~Rev.~D, {\bf 15}, 307 (1985);\\
M.~Razavy, Phys.~Rev.~A, {\bf 43}, 3486 (1993).
\bibitem{pra97b} A.~Mostafazadeh, Phys.~Rev.~A {\bf 55}, 4084 (1997).
\bibitem{pla97b} A.~Mostafazadeh, Phys.~Lett.~A {\bf 232}, 395 (1997).
\bibitem{ga-go-ro-th} G.~Giavarini, E.~Gozzi, D.~Rohrlich, and 
W.~D.~Thacker, Phys.~Lett.~A {\bf 138}, 235 (1989).
\bibitem{optics} R.~Bhandari, Phys.~Rep.~{\bf 281}, 1 (1997).
\bibitem{jpa98b} A.~Mostafazadeh, J.~Phys.\ A: Math.\ Gen.\ {\bf 31}, 
7829 (1998).
\end{thebibliography}
\end{document}